\input{aipcheck}

\documentclass[
    ,final            
  ]
  {aipproc}

\layoutstyle{8x11single}

\graphicspath{{fig/}}

\begin{document}

\title{Non-congruence of liquid-gas phase
transition of asymmetric nuclear matter}


\author{Toshiki Maruyama}{
  address={Japan Atomic Energy Agency,
Shirakata Shirane 2-4, Tokai, Ibaraki, 319-1195 Japan}
}

\author{Toshitaka Tatsumi}{
  address={Department of Physics, Kyoto University,
Kyoto, 606-8502 Japan}
}

\begin{abstract}

We first explore the liquid-gas mixed phase in a bulk calculation, 
where two phases coexist without the geometrical structures. 
In the case of symmetric nuclear matter, the system behaves congruently, and   
the Maxwell construction becomes relevant. 
For asymmetric nuclear matter, on the other hand, 
the phase equilibrium is no more attained by the Maxwell construction 
since the liquid and gas phases are non-congruent; 
the particle fractions become completely different with each other.
One of the origins of such non-congruence is attributed to
the large symmetry energy. 

Subsequently we explore the charge-neutral nuclear matter 
with electrons by fully applying the Gibbs conditions to 
figure out the geometrical (pasta) structures in the liquid-gas mixed phase.  
We emphasize the effects of the surface tension and the Coulomb
interaction on the pasta structures. 
We also discuss the thermal effects on the pasta structures.
\end{abstract}

\maketitle


\section{Introduction}

The liquid-gas (LG) phase transition and its relevant equation of state (EOS) is 
one of the most important issues in nuclear physics and astrophysics. 
In the collapsing stage of supernovae and the crust region of 
compact stars,
low-density nuclear matter with non-uniform structures has been expected. 
Such structured matter can be regarded as a mixed phase during 
the LG phase transition. 
The mixed phase in thermal equilibrium is often obtained by simply applying
the Maxwell construction or more carefully by imposing the Gibbs conditions. 
However, the Maxwell construction can be used in the case of 
congruent transition \cite{ios},
i.e., the coexisting two phases have the same particle fractions 
with different total densities. 
A familiar example is the isotherm of boiling water which EOS exhibits a
constant pressure in the mixed phase.

On the other hand, 
for asymmetric nuclear matter or charge-neutral nuclear matter 
which consists of proton, neutron and electron, 
the phase transition is generally non-congruent.
The particle fractions take different values in each phase. 
The Gibbs conditions (equilibrium of chemical potentials, temperature and pressure 
between two phases) then give EOS different from that of 
the Maxwell construction.

\section{Bulk calculation with the relativistic mean-field model}

First, we study the phase coexistence in low-density nuclear matter 
without considering any geometrical structure.
We use the relativistic mean-field (RMF) model for the interaction.
In the RMF model, baryons interact via exchange of 
$\sigma$, $\omega$ and $\rho$ mesons. 
All the fields of baryons and mesons are introduced 
in a Lorentz-invariant way.
It is not only relatively simple for 
numerical calculations, but also sufficiently
realistic to reproduce bulk properties of finite nuclei
as well as the saturation properties of nuclear matter \cite{maru05,marurev}.%
One characteristic of our framework is that
the Coulomb interaction is properly included in the 
equations of motion for nucleons and electrons and for meson mean fields.
Thus the baryon and electron density profiles, as well as the meson
mean fields, are determined in a fully
self-consistent way with the Coulomb interaction.

\begin{figure}[h]
\centerline{
\includegraphics[width=.68\textwidth]{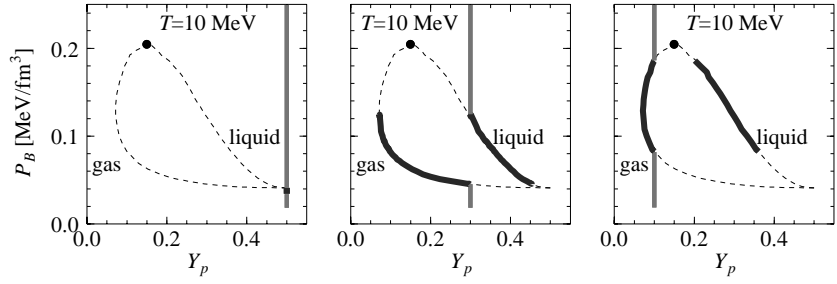}
}
\caption{
Binodal curves (dashed curves) in $Y_p$-$P_B$ plane at $T=10$ MeV.
Filled circles show the critical points.
Vertical lines and thick curves on the binodal curves indicate 
trajectories of isothermal compression.
}
\label{figFracPa}
\end{figure}

The equilibrium configuration of the liquid (L) and gas (G) phases is obtained
by solving the Gibbs conditions as
$T^G=T^L,\quad P^G=P^L,\quad \mu_i^G=\mu_i^L (i=n,p),$
where $\mu_i$ denotes the chemical potential of ingredient $i$.
for given temperature $T$ and baryon partial pressure $P_B$, 
we can find generally two values of $Y_p$ to satisfy the Gibbs conditions.
However, there are the maximum and minimum values of $P_B$ which can give two values of $Y_p$.
Consequently, changing the pressure we can draw a closed curve (binodal curve) 
by connecting these values in $Y_p$-$P_B$ plane.
In Fig. \ref{figFracPa}, the dashed line shows the phase-coexistence (binodal) curves at $T=10$ MeV.
Density at the region right from the binodal curve (most symmetric)
is generally higher than the left (neutron rich) region 
due to the symmetry energy. 
Therefore we call the left region a gas and the right region liquid. 
The inner region of binodal curve is thermodynamically forbidden region.
If we put asymmetric nuclear matter with $Y_p=0.3$ with very low pressure
(i.e., a gas) and start to compress it, the trajectory in 
the $Y_p$-$P_B$ plane looks as a vertical line in the central panel.
At some point, the line encounters the binodal curve.
As the system cannot enter the inner region,
the system splits into two phases with the same $P_B$ and different $Y_p$,
which corresponds the appearance of mixed phase.
When we continue compression of the system, 
there appears the phase separation of gas and liquid 
with the various volume fraction like an amorphous state. Eventually $Y_p$ of the right curve 
equals to that of the total system $Y_p=0.3$, which 
 means that the phase on the right side, i.e.\ liquid, dominates
the whole system. 
After that the compression leads the vertical line to go upward.
The above change represents a {\it non-congruent} phase transition 
from the gas to liquid phase.

In the case of symmetric nuclear matter $Y_p=0.5$, which is shown 
in the left panel, the vertical line encounters the binodal curve
at one point.
Therefore the phase transition from gas to liquid occurs with 
a constant value of $Y_p=0.5$.
This means the phase transition is {\it congruent}. 
The pressure is also constant during the phase transition,
which can be obtained by the Maxwell construction.
In the case of $Y_p=0.1$ (neutron-rich),
the behavior at lower pressure is the same as the above two cases.
However, it should be interesting to see the curve on the left side encounters the value of the 
total system 0.1 twice, which means 
the transition proceeds from gas to the mixed phase and back to gas again.
This is called ``retrograde condensation'', which was observed
in the mixture of organic solvent \cite{lan}. 
Note that the gas state at large pressure is not a simple gas 
but the supercritical one.

\begin{figure}[h]
\centerline{
\includegraphics[width=.36\textwidth]{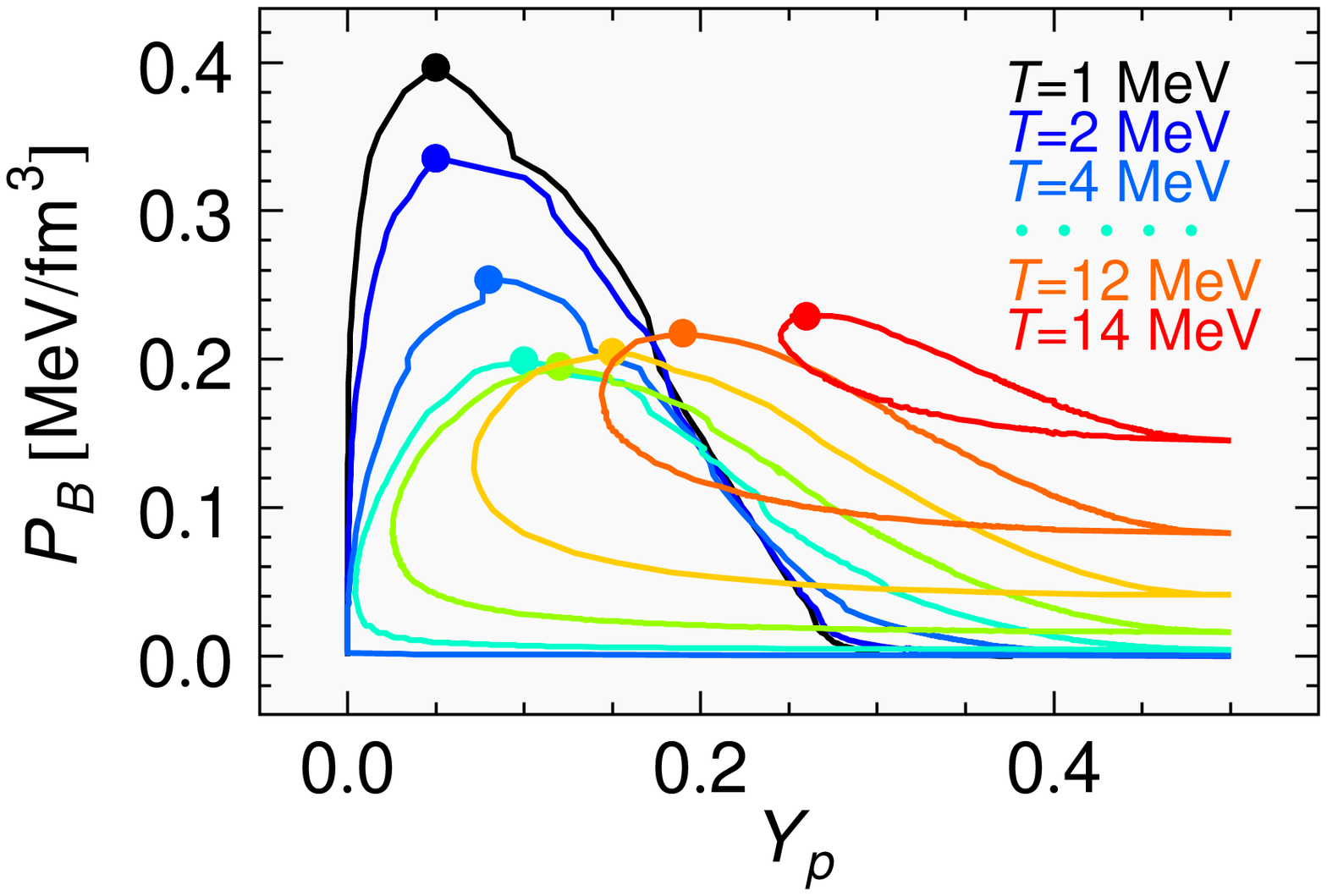}
\includegraphics[width=.36\textwidth]{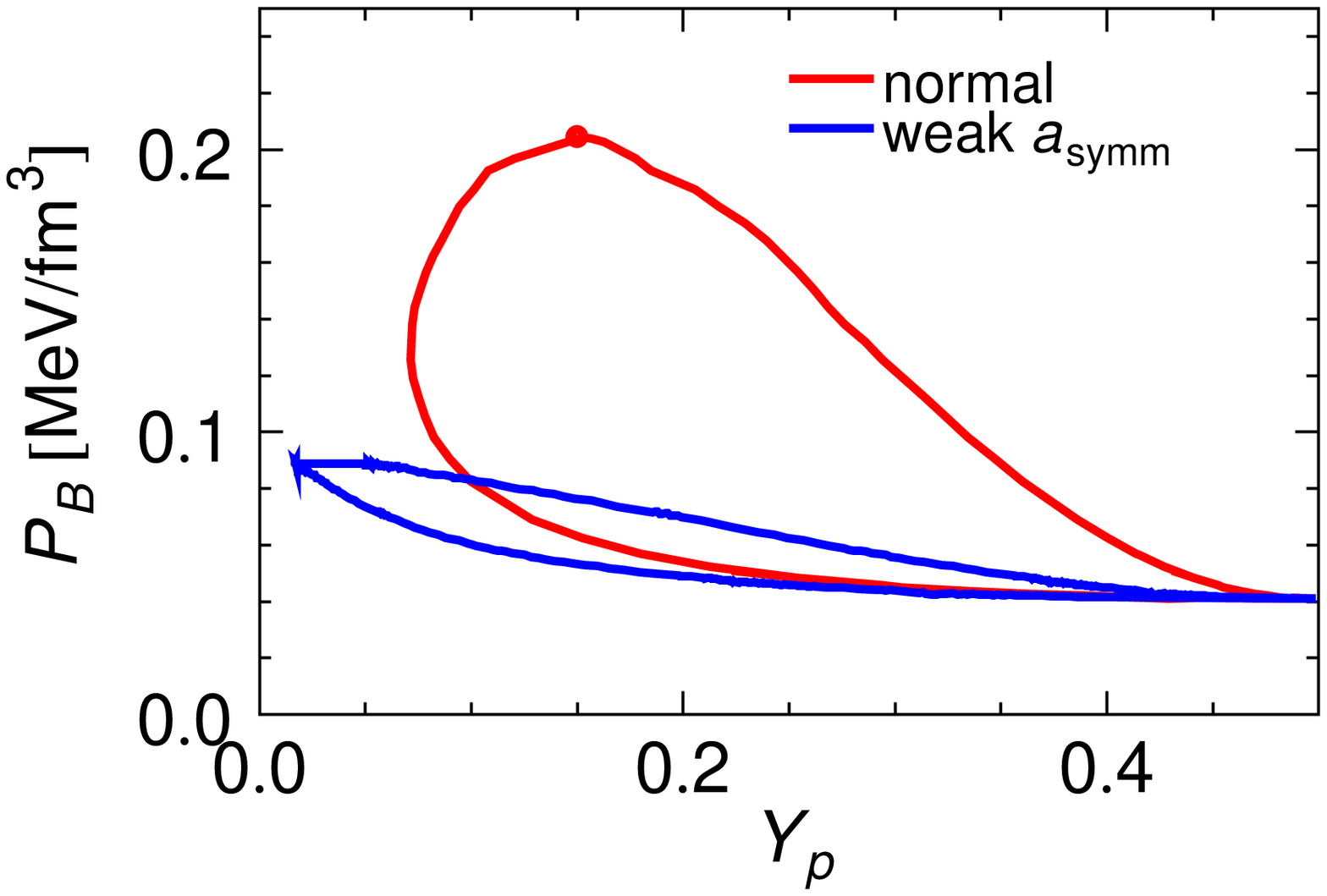}
}
\caption{
Temperature dependence (left)
and 
symmetry-energy dependence (right) 
 of the binodal curves in $Y_p$-$P_B$ plane.
Filled circles show the critical points.
\vspace{-1mm}}
\label{YP}
\end{figure}

The left panel of Fig.\ \ref{YP} shows the binodal curves 
for various temperatures. 
At low temperatures, the forbidden region surrounded by the
binodal curve is wide.
Below 4 MeV, the left part of the binodal curve touches to
the origin of the vertical axis $Y_p=0$, which means the gas phase
is vacant or pure neutron matter. 
At higher temperatures, the region surrounded by the binodal curve
becomes narrow, which indicates the 
enhancement of the congruence.

The non-congruence of asymmetric nuclear matter is
coming from the strong attraction between proton and
neutron (symmetry potential) which makes 
the particle fraction of the liquid phase more symmetric 
than the total system.
In fact, if we employ a weak symmetry potential in 
the bulk calculation, the binodal curve will collapse and 
the congruence will be enhanced (see the right panel of Fig.\ \ref{YP}).

\section{Property of nuclear matter with the pasta structures}

\begin{figure}[h]
\centerline{
\includegraphics[width=.38\textwidth]{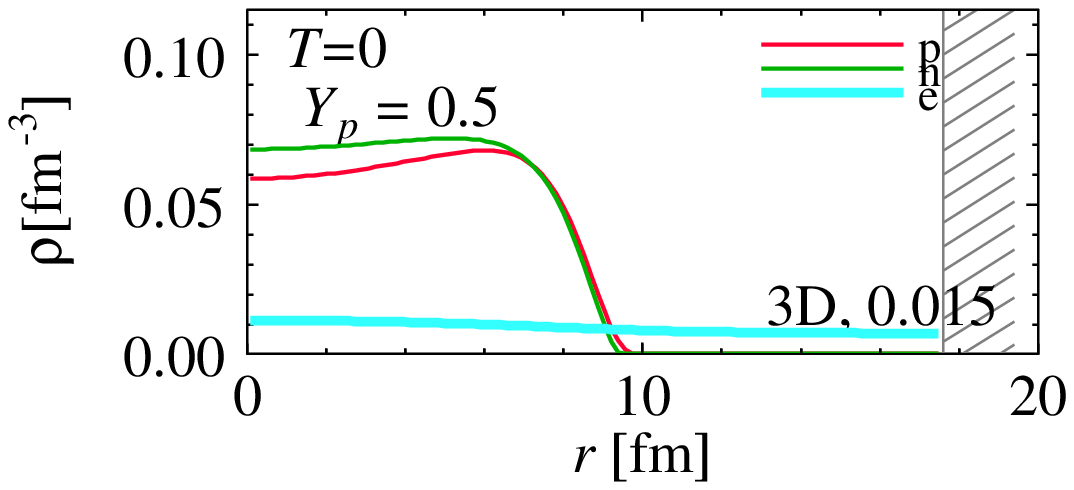}
\includegraphics[width=.38\textwidth]{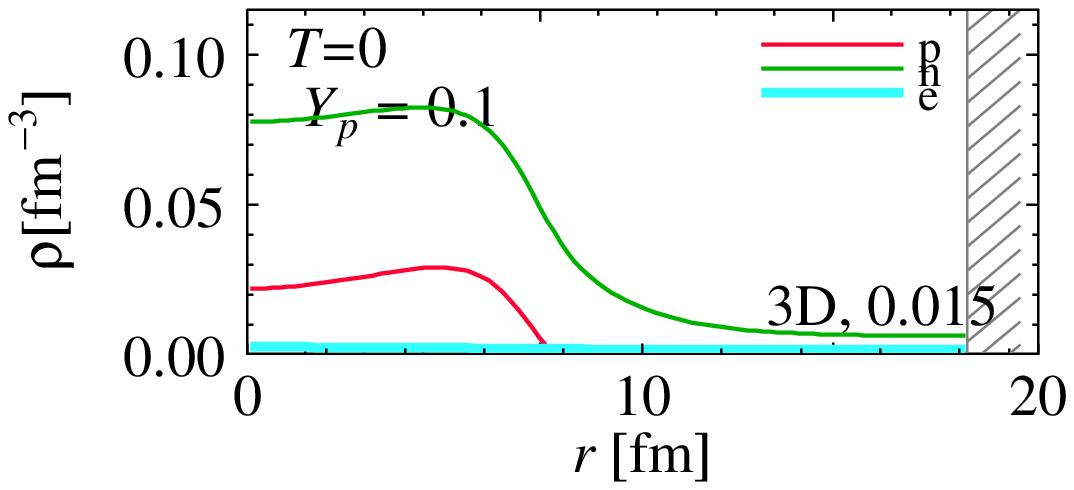}
}
\caption{
Density profiles of symmetric (left) and asymmetric (right) nuclear matter. 
}
\label{figProf}
\end{figure}

Next let us discuss the property of inhomogeneous nuclear matter.
It is expected that inhomogeneous nuclear matter exhibits so-called
the pasta structures at subnuclear densities \cite{Rav83,Has84}.
The origin of the pasta structures comes from the balance between 
the Coulomb energy and the surface energy of matter 
which are dependent on the geometry such as
droplet, rod, slab, tube, bubble, and uniform. 
To obtain matter with the pasta structures, under the local-density approximation, 
we solve the equations of motion for
the meson mean-fields which are dependent on the position, 
and the relation between chemical potentials and fermion densities.  
Details are explained in Refs.\ \cite{maru05,marurev,maruPTPS}.

Some examples of the density profiles (droplets) are shown in 
Fig.\ \ref{figProf}.
In the case of symmetric nuclear matter, density profiles
of proton and neutron are almost equal, showing the congruence
of the phase transition. 
On the other hand, asymmetric nuclear matter is non-congruent,
 accompanying the phase separation 
with different proton fractions in the liquid and gas phases.

\begin{figure}[h]
\centerline{
\includegraphics[width=.38\textwidth]{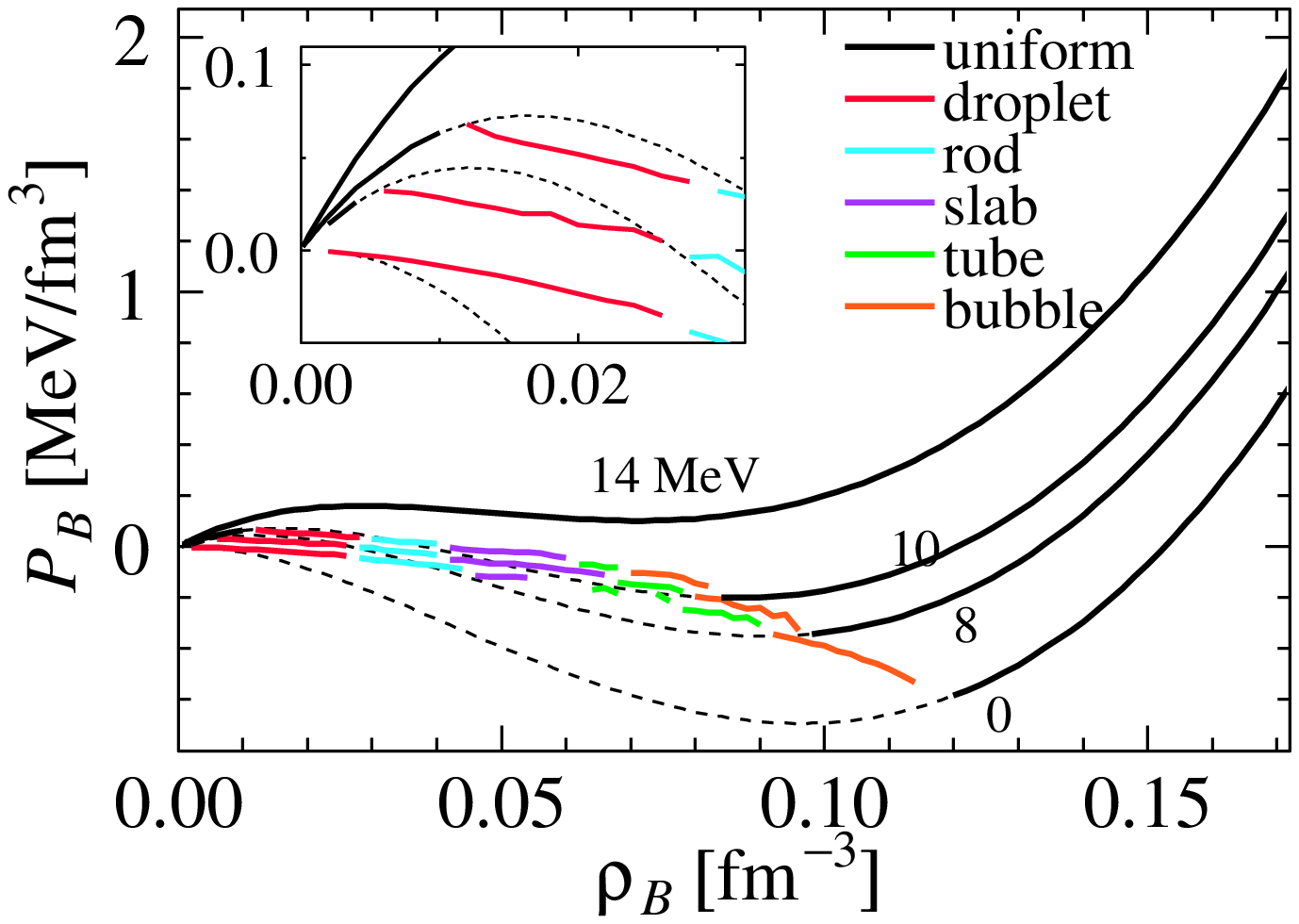}
\includegraphics[width=.38\textwidth]{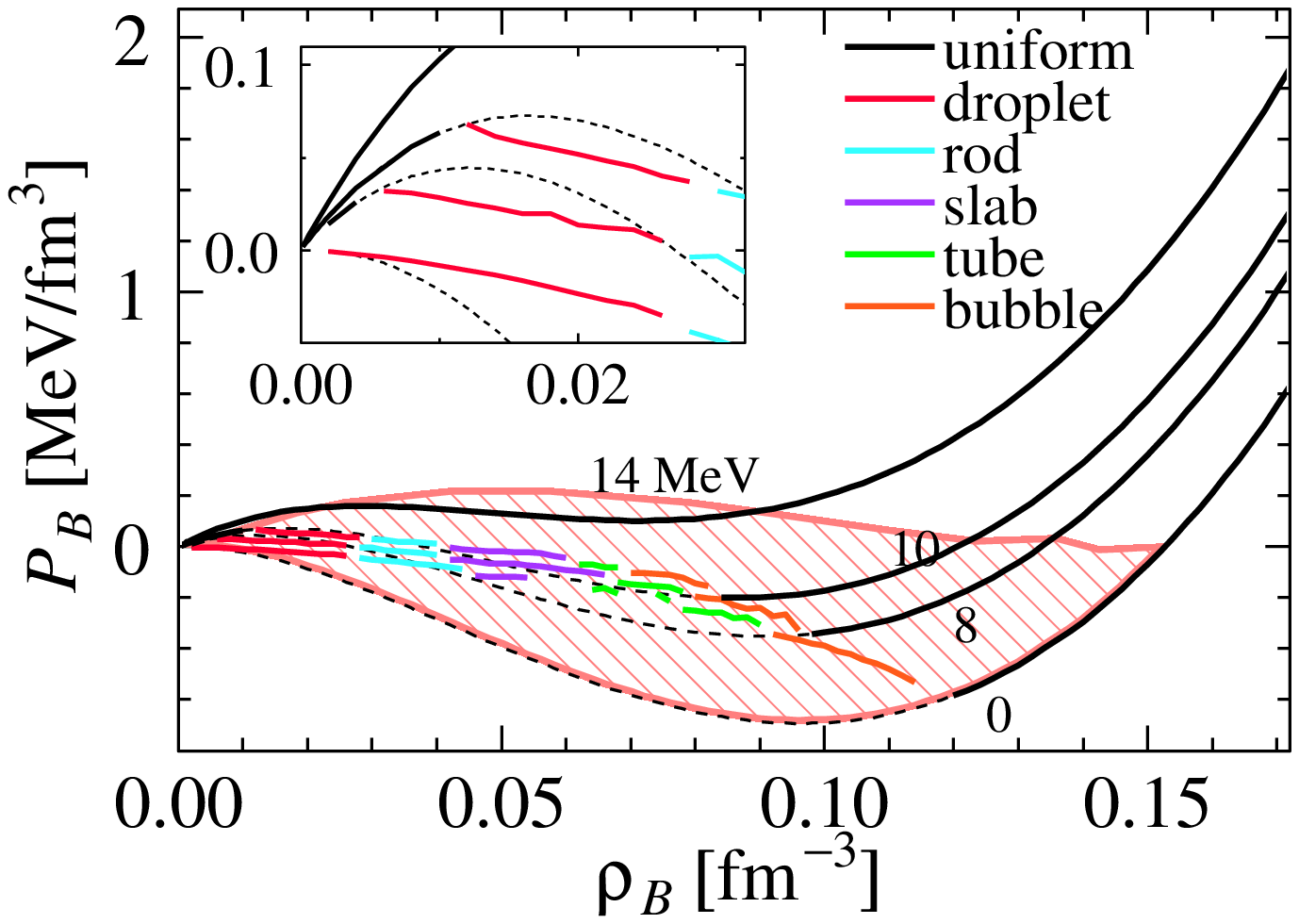}
}
\caption{
Left: EOS of symmetric matter with the pasta structures.
Right: Hatched area shows the binodal region. 
}
\label{figEOSpasta}
\end{figure}

The resulting EOS of symmetric nuclear matter at various temperatures
are presented in the left panel of Fig.\ \ref{figEOSpasta}. 
Colored curves indicate appearance of various pasta structures
while smooth black curves uniform matter for comparison.
In the right panel indicated with hatch is the binodal region.
We can see that the region where the pasta structures appear is
inside the binodal region.
Near the upper (higher density) boundary  of the binodal region,
the appearance of the pasta structures is suppressed 
by the Coulomb and surface energies
which are characteristic of the structured mixed phase.



\bibliographystyle{aipproc}   

\bibliography{sample}

\IfFileExists{\jobname.bbl}{}
 {\typeout{}
  \typeout{******************************************}
  \typeout{** Please run "bibtex \jobname" to obtain}
  \typeout{** the bibliography and then re-run LaTeX}
  \typeout{** twice to fix the references!}
  \typeout{******************************************}
  \typeout{}
 }


\end{document}